\documentclass[a4paper]{jpconf}
\usepackage{graphicx}
\begin{document}
\title{ Pseudospin Dynamical Symmetry in Nuclei}

\author{Joseph N Ginocchio}

\address{ MS 283, Los Alamos National Laboratory, Los Alamos, NM, 87545,USA}

\ead{gino@lanl.gov}

\begin{abstract}
Pseudospin symmetry has been useful in understanding atomic nuclei. We review the arguments that this symmetry is a relativistic symmetry. The condition for this symmetry is that the sum of the vector and scalar potentials in the Dirac Hamiltonian is a constant. We give the generators of pseudospin symmetry. We review some of the predictions that follow from this insight into the relativistic origins of pseudospin symmetry. Since in nuclei the sum of the scalar and vector potentials is not zero but is small, we discuss preliminary investigations into the conditions on the potentials to produce partial dynamic pseudospin symmetry. Finally we show that approximate pseudospin symmetry in nuclei predicts approximate spin symmetry in anti-nucleon scattering from nuclei.

\end{abstract}

\section{Introduction}

Fifty years ago Aldo and I first met at Rutgers University. We both were starting our first post-doctoral appointments. We have remained in touch primarily through the 10 seminars in nuclear physics that he has organized in different venues in and near the Bay of Naples for the last 28 years. These seminars have been enormously popular and successful. I have enjoyed knowing Aldo and Renata these last fifty years.
 
When Aldo and I started our careers nuclear physics was a very exciting field of research. About five years after our post-doctoral appointments at Rutgers, a quasi-degeneracy in the single nucleon states of spherical nuclei with quantum numbers $(n\ \ell_{j},n^{\prime}{\ell}^{\prime}_{j^{\prime}})$ was discovered  \cite{arima,hecht}, where $n^{\prime}=n-1, \ell^{\prime}=\ell+2,j^{\prime}=j+{1}$ and  $n,\ell, j $ are the radial, orbital angular momentum, and total angular momentum quantum numbers, respectively. 
These quasi-degeneracies persist in recent measurements in nuclei far from stability \cite {sorlin08}. The authors realized that, if they define the average of the orbital angular momenta as a pseudo-orbital angular momentum (${\tilde {\ell}}$) and then couple a pseudospin (${\tilde {s}} = {1 \over 2}$) to the pseudo-orbital angular momentum, they will get the total angular momenta (${j}= {\tilde {\ell}}\pm  {1 \over 2} $). For example, for the  $(1s_ {1\over 2},0d_ {3\over 2})$ orbits, ${\tilde {\ell}}$ = 1, which gives the total angular momenta $j= {1\over 2}, {3\over 2}$. Subsequently pseudospin doublets in deformed nuclei were discovered \cite {bohr}. Pseudospin symmetry was later revealed to be a symmetry of the Dirac Hamiltonian \cite{gino97,gino05}.

\section{Symmetries of the Dirac Hamiltonian}
The Dirac Hamiltonian with a Lorentz scalar potential, $V_S(\vec r)$, and a potential which is the fourth component of a  Lorentz vector potential, $V_V(\vec r)$, is 
\begin{equation}
H ={\vec \alpha}\cdot {\vec p}
+ \beta (V_S(\vec r) + M) + V_V(\vec r),
\label {dirac}
\end{equation}
where ${\vec \alpha}$, $\beta$ are the Dirac matrices,  ${\vec p}$ is the momentum, $M$ is the mass, ${\vec r}$ is the radial coordinate, and the velocity of light is set equal to unity, c =1. 

\subsection{Spin Symmetry: A Symmetry of the Dirac Hamiltonian}
The Dirac Hamiltonian has spin symmetry when the difference of the vector and scalar potentials in the Dirac Hamiltonian is a constant, $V_S(\vec r) - V_V(\vec r)= C_{s}$ \cite {bell75}. Hadrons \cite {page01} and anti-nucleons in a nuclear environment have spin symmetry \cite {gino99}. These are relativistic systems and normally, in such systems, we would expect large spin-orbit splittings, but, in this limit, spin doublets are degenerate.
The generators for this SU(2) spin algebra, 
${{\vec S}}$,
which commute with the
Dirac Hamiltonian with any potential $V(\vec r)$, spherical or deformed, $[\,H\,,\, {\vec S}\,] = 0$, 
are given by \cite {gino98}
\begin{equation}
{\vec {S} }=
\left (
\begin{array}{cc}
  {\vec s} 
  & 0 \\
0 &U_p\ {\vec  s}\ U_p
\end{array}
\right ), \ 
\label{gen}
\end{equation}
where
$ {\vec s} = {\vec \sigma}/2$ are the usual spin generators,
${\vec \sigma}$ the Pauli matrices, 
and $U_p = \, {\mbox{\boldmath ${\vec \sigma}\cdot {\hat p}$} }$ is the
helicity unitary operator introduced in \cite {blo95} and ${\hat p}={\vec p\over p}$, the unit momentum. The generators are four by four matrices as appropriate for the Dirac Hamiltonian. 

\subsection{Pseudospin Symmetry: A Symmetry of the Dirac Hamiltonian}
Another SU(2) symmetry of the Dirac Hamiltonian occurs when the sum of the vector and scalar potentials in the Dirac Hamiltonian is a constant, $V_S(\vec r)$ + $V_V(\vec r)= C_{ps}$  \cite {bell75}. The generators for this SU(2) algebra 
${\vec {\tilde S} }$,
which commute with the
Dirac Hamiltonian with any potential $V(\vec r)$, spherical or deformed, $[\,H\,,\, {\vec {\tilde S} }\,] = 0$, 
are given by \cite {gino98}
\begin{equation}
{\vec {\tilde S} }=
\left (
\begin{array}{cc}
U_p\ {\vec  s}\ U_p
  & 0 \\
0 &  {\vec s} 
\end{array}
\right ). \ 
\label{genps}
\end{equation}

This symmetry was shown to be pseudospin symmetry \cite {gino97}. The eigenfunctions of the Dirac Hamiltonian in this limit will have degenerate doublets of states, one of which has pseudospin aligned and the other with pseudospin unaligned. 
The ``upper" matrix of the pseudospin generators  in Eq (\ref {genps}), $U_p\ {\vec  s}\ U_p$,  have the spin intertwined with the momentum which enables the generators to connect the states in the doublet, which differ by two units of angular momentum.
In finite nuclei both potentials go to zero at infinite distances. Hence $C_{ps}=0$ and the scalar potential will be equal in magnitude but opposite in sign to the vector potential in the pseudospin limit. The approximate equality in the magnitude of the vector and scalar fields in nuclei and their opposite sign have been confirmed in relativistic mean field theories \cite  {gino05} and in QCD sum rules \cite {cohen,gino05}.
\section{Consequences of  Relativistic Pseudospin Symmetry}

One immediate consequence of pseudospin symmetry as a relativistic symmetry is that the ``lower" matrix of the pseudospin generators in Eq (\ref {genps}), $\vec  s$, does not change the radial wavefunction of the ``lower" component of the Dirac eigenfunctions. Hence this symmetry predicts that the radial wavefunctions of the ``lower" component are the same for the two states in the doublet. Previous to this discovery many relativistic mean field calculations of nuclear properties had been made. Hence this prediction was tested with existing calculations and, indeed, these wavefunctions are very similar for both spherical \cite {ginoPRC98,ring} and deformed nuclei \cite {arima02,gino04}. Because of the momentum dependence of the ``upper" matrix of the generators the relationship between the ``upper" components involves a differential equation and these have also been tested in spherical \cite {gino02} and deformed nuclei \cite {gino04} with success.

Magnetic dipole and Gamow-Teller transitions between the two single-nucleon states in pseudospin doublets are forbidden non-relativistically because the states differ by two units of angular momentum. However, they are not forbidden relativistically which means that they are proportional to the lower component of the Dirac eigenfunction. This leads to  a condition between the magnetic moments of the states and the magnetic dipole transtion between them because  the radial amplitudes of the lower components of the two states in a pseudospin doublet are equal. Therefore  the magnetic dipole transition between the two states in the doublet can be predicted if the
magnetic moments of the states are known \cite{ginoPRC99,gino05}. Likewise pseudospin symmetry also predicts Gamow-Teller transitions between a state in a parent nucleus to the partner state in the daughter nucleus if the Gamow-Teller transition to the same states in the parent and daughter nucleus is known. We do not have space to discuss these relationships in detail but one 
example occurs in
the mirror nuclei $^{39}_{19}$K$_{20}$ and $^{39}_{20}$Ca$_{19}$. The ground state and first excited state
of $^{39}_{19}$K$_{20}$ are interpreted as a $0d_{3/2}$ and
$1s_{1/2}$ proton hole respectively, while the ground
state and first excited state of $^{39}_{20}$Ca$_{19}$ are
interpreted as a $0d_{3/2}$ and $1s_{1/2}$ neutron hole
respectively. These states are members of the ${{\tilde n}_r } = 1,
{\tilde {\ell}} =1$
pseudospin doublet. Using the magnetic moment of $^{39}$Ca a
transition rate is calculated which is only about 37 $ \% $ larger than the measured value.
However, the two states in the doublet are not pure single-particle
states. A modification of these relations has been derived which take into account the fact
that these states are not pure single
particle states  \cite{gino00,gino05}. The modified relations give a transition rate that
agrees with the measured value to within
experimental error.
Again using the mass 39 nuclei, the Gamow - Teller transitions from the ground state of
$^{39}$Ca to the ground and
first excited state of $^{39}$K are known, which is enough
information to predict the transition from the ground state to the excited state.
In the non-relativistic shell model an effective tensor
term $g_{eff}\ [ Y_2
\sigma]^{(1)}$ is added to the magnetic dipole operator and the Gamow-Teller operator to produce a
transition, where $g_{eff}$ is a calculated effective
coupling constant. However, the 
magnetic dipole transition calculated
between the same states is an order of magnitude lower than the
experimental transition \cite {ian94} although the calculated Gamow-Teller  agrees with the experimental
value within the limits of experimental
and theoretical uncertainty. This inconsistency has been a
puzzle for the non-relativistic shell model.
On the other hand the relativistic
single-nucleon model gives a consistent description
of both of these transitions. A global prediction of magnetic dipole transitions throughout the periodic table has had reasonable success as well  \cite{gino00,gino05}. However, a global prediction of Gamow-Teller transitions have not been done yet. Pseuodspin symmetry can also be used to relate quadrupole transitions between multiplets \cite{gino05}.
\section{Pseudospin as a Partial Dynamical Symmetry}

In the invariant pseudospin symmetry limit there are no bound states for nucleons above the Dirac sea \cite{gino05}. Hence pseudospin symmetry can never be an invariant symmetry. Experimental evidence and theoretical calculations suggest that the symmetry is better preserved near the Fermi sea \cite{gino05}. Also the ``intruder" states, states that have zero nodes and have spin aligned with the orbital angular momentum, do not have bound pseudospin partners and therefore the intruder states definitely do not have pseudospin symmetry \cite{gino01}. For this reason it is of interest to investigate whether pseudospin symmetry is a partial dynamical symmetry \cite{ami96} . A nucleus with partial dynamical system is one in which some of the states in a nucleus may have the symmetry, while others do not. The question is then: ``Are there Dirac Hamiltonians that have a partial dynamical pseudospin symmetry?".  

We have started to address this question. We shall first consider spherically symmetric Dirac Hamiltonians. For such Hamiltonians the alignment of the spin with the orbital angular momentum is conserved. The conserved quantum number is $\kappa$ and 

\begin{eqnarray}
\kappa &= &(j + 1/2), \   j=\ell - 1/2 = \tilde{\ell} + 1/2    \nonumber \ \
\end{eqnarray}
\begin{eqnarray}
 \kappa  &=& -(j + 1/2), \   j= \ell +1/2 = \tilde{\ell} - 1/2.
\label {kappa}
\end{eqnarray}
The two states that are pseudospin doublets we label as $\kappa_a$ (pseudospin aligned) and $ \kappa_u$  (pseudospin unaligned), where

\begin{eqnarray}
\kappa_a &= & \tilde{\ell}  + 1, \  j= \tilde{\ell} + 1/2    \nonumber \ \
\end{eqnarray}
\begin{eqnarray}
 \kappa_u  &=& - \tilde{\ell} , \   j= \tilde{\ell} - 1/2.
\label {kappad}
\end{eqnarray}
so that $\kappa_a +  \kappa_u =1$ for the two states in a pseudospin doublet. Denoting $G_{\kappa,j}(r)/r$ as the radial wavefunction of the upper component of the Dirac eigenfunction, the condition that these two states are members of a pseudospin doublet is
\begin{equation}
y_{\tilde \ell}(r)=G'_{\kappa_a,j}(r) +{ \kappa_a\over r} \ G_{\kappa_a,j}(r)=G'_{ \kappa_u,j}(r) +{\kappa_u\over r} \ G_{\kappa_u,j}(r).
\label{y}
\end{equation}
The function $y_{\tilde \ell}(r)$ depends on ${\tilde \ell}$ but not on the two $\kappa$'s of the states in the doublet. The Dirac equation implies a second order differential equation for  $G_{\kappa,j}(r)$ which in turn implies the following second order differential equation for $y_{\tilde \ell}(r)$
\begin{equation}
y''_{\tilde \ell}(r) + C_{\kappa}^{(1)} y'_{\tilde \ell}(r)+ C_{\kappa}^{(0)} y_{\tilde \ell}(r)=0,
\label{ydiff}
\end{equation}
where
\begin{eqnarray}
C_{\kappa}^{(1)} = -\left(2{A'_{\kappa}(r) \over A_{\kappa}(r)} + {B'_{\kappa}(r) \over B_{\kappa}(r)}\right),    \nonumber
\end{eqnarray}
\begin{eqnarray}
C_{\kappa}^{(0)} =-{{{\tilde\ell} ({\tilde\ell}+1)}\over r}-{A''_{\kappa}(r) \over A_{\kappa}(r)}+ 2\left({A'_{\kappa}(r) \over A_{\kappa}(r)}\right )^2+ \left({\kappa \over r} + {A'_{\kappa}(r) \over A_{\kappa}(r)}\right ){B'_{\kappa}(r) \over B_{\kappa}(r)}+  A_{\kappa}(r)B_{\kappa}(r),
\label {C}
\end{eqnarray}
and 
\begin{eqnarray}
A_{\kappa}(r)=E_{\kappa}+M +V_S( r)  - V_V( r), \ B_{\kappa}(r)=E_{\kappa}-M -V_S( r)  - V_V( r),
\end{eqnarray}
where $E_{\kappa}$ is the eigenenergy of the Dirac eigenfunction. The condition then for two states to be pseudospin doublets is $C_{\kappa_a}^{(i)}=C_{\kappa_u}^{(i)}$ where $i=0,1$. Not every eigenstate will  satisfy these conditions. We are investigating the possible solutions of these conditions.

Another approach is to require that the commutator of the Dirac Hamiltonian with the pseudospin generators not be identically zero but zero only when acting on the states which have good pseudospin symmetry; i.e.,
\begin{eqnarray}
 [\,H\,,\, {\vec {\tilde S} }]\ \Psi_{\kappa,j}(\vec r)  = 0,  
 \label {Com}
 \end{eqnarray}
for both the states in the doublet, $\kappa=\kappa_a, \kappa_u$, and where $\Psi_{\kappa,j}(\vec r)$ is the two component Dirac eigenfunction. This leads to a condition only on the upper component 
\begin{eqnarray}
 [\,V_S( r)  + V_V( r)\,,\, U_p\ {\hat {p} }]\ G_{{\kappa_{a,u}},j}(r)/r \left[Y^{({\tilde\ell} \pm 1)}\chi\right]^{j={\tilde\ell}\pm {1\over2}}= 0, 
 \label {Comu}
 \end{eqnarray}
 where $Y^{({\ell} )}$ is the spherical harmonic of rank $\ell$ and $\chi$ is the spin function and $[ \ ]^{j}$ means coupled to angular momentum $j$. These conditions are being studied to see if there are potentials which satisfy these conditions.
\section{Anti-nucleon in a Nuclear Environment}
Charge conjugation changes a nucleon into an anti-nucleon. Under charge conjugation the scalar potential of a nucleon remains invariant while the its vector potential changes sign. Hence the pseudospin condition $V_S(\vec r) + V_V(\vec r)\approx C_{ps}$ becomes the spin condition ${\bar V}_S(\vec r) - {\bar V}_V(\vec r)\approx C_{ps}=\bar{C}_s$ where $ {\bar V}_{S,V}(\vec r)$ are the scalar and vector potentials of an anti-nucleon in a nuclear environment respectively. Hence approximate pseudospin for nucleons predicts approximate spin symmetry for anti-nucleons in a nuclear environment.  Since the potentials are also very deep we would expect approximate $U(3)$ symmetry as well \cite {ginoprl05}. 

Of course the anti-nucleons scattered from nuclei experience a strong annihilation potential while nucleons scattered from nuclei do not. However, since the nucleons do not feel an annihilation potential, this means that the sum of the scalar and vector annihilation potentials for nucleons must be zero which means that the difference of the anti-nucleon annihilation potentials is zero as well. Hence the anti-nucleon annihilation potentials conserve spin symmetry as well.

The sparse data on the scattering of polarized anti-protons on nuclei supports the prediction that spin symmetry is approximately conserved \cite {martin}. Perhaps more extensive data on the scattering of polarized anti-protons on nuclei will be forthcoming from GSI.
 \section{Conclusions}
Most features of nuclei can be understood within the non-relativistic shell model. Therefore it comes as a surprise that  pseudospin symmetry in nuclei is a relativistic symmetry.  However, it can never be an invariant symmetry since no nucleon bound states exist when the sum of the scalar and vector potentials is a constant. Investigations have begun to determine if certain potentials produce a  partial dynamic symmetry. 

Approximate pseudospin symmetry in nuclei implies that spin symmetry will be approximately conserved in anti-nucleon scattering from nuclei. The new anti-proton facility at GSI could test this prediction by measuring the polarization of anti-protons scattered from nuclei. This prediction follows simply from charge conjugation and, if confirmed, will be another manifestation of the relativistic origin of pseudospin symmetry. 

In addition, since QCD sum rules predict that the sum of the scalar and vector potentials is small in nuclear matter, perhaps there is a more fundamental rationale for pseudospin symmetry in terms of quark dynamics.\\

\noindent{\bf Acknowledgments}

\noindent{The author would like to thank Prof. Ami Leviatan for his collaboration on the preliminary research on pseudospin symmetry as a partial dynamic symmetry discussed in Section 4. This research was supported by the US Department of Energy, Contract No. W-7405-ENG-36.}
\section{References}

\end{document}